\documentclass[aps,pra,preprint,amsmath,amssymb,superscriptaddress,showpacs,floatfix]{revtex4-1}
\usepackage{latexsym}
\usepackage{graphicx}
\newcommand\beq{\begin{equation}}
\newcommand\eeq{\end{equation}}
\newcommand\bea{\begin{eqnarray}}
\newcommand\eea{\end{eqnarray}}
\newcommand\non{\nonumber}
\begin{document}

\title{A Study of Topological Quantum Phase Transition
and Majorana Localization Length for the Interacting Helical Liquid
System}
\author{ Dayasindhu Dey}
\affiliation{S. N. Bose National Centre for Basic Sciences, Block - JD, Sector - III, Salt Lake, Kolkata - 700098, India}

\author{Sudip Kumar Saha}
\affiliation{S. N. Bose National Centre for Basic Sciences, Block - JD, Sector - III, Salt Lake, Kolkata - 700098, India}

\author{P. Singha Deo}
\affiliation{S. N. Bose National Centre for Basic Sciences, Block - JD, Sector - III, Salt Lake, Kolkata - 700098, India}

\author{Manoranjan Kumar}
\affiliation{S. N. Bose National Centre for Basic Sciences, Block - JD, Sector - III, Salt Lake, Kolkata - 700098, India}

\author{Sujit Sarkar}
\affiliation{Poornaprajna Institute of Scientific Research, 4 Sadashivanagar, Bangalore 5600 80, India}

\date{\today}

\begin{abstract}
We consider a helical spin liquid system which shows majorana fermion modes 
at the edge. The interaction between the quasiparticles in this system induces 
phase transition, Majorana-Ising transition. We comply the density matrix 
renormalization group method to study this phase 
transition for the entire regime of the parameter space. 
We observe the presence of topological quantum phase transition for 
repulsive interaction, however this phase is more stable for the attractive 
interaction. The length scale dependent study shows many new and 
important results and we show explicitly that the major contribution to the excitation comes 
from the edge of the system when the system is in the topological state. We also
show the dependence of Majorana localization length for various values of 
chemical potential. 
\end{abstract}

\keywords{Topological Insulator, Helical Liquid, Topological Quantum Phase Transition} 

\maketitle
\section{Introduction}
The physics of Majorana fermion is the subject of intense research 
in quantum many-body systems over  a  decade~\cite{wil,majo1,berni} 
which appears in the topologically ordered state. The presence  of 
this   particle   is   not   only     searched in neutrino physics, 
supersymmetry and dark matter, but it also appears as an  emergent 
particle  in  condensed  matter  systems,  such as one-dimensional 
superconductor~\cite{ivanov,kraus,wimmer,sato}, 
semiconductor quantum wire~\cite{mourik,deng,rokin,das,finc,jafari}, 
proximity induced topological 
superconductor~\cite{kitaev,fu1,fu2,ali1,ali,sau,potter,ali2,fukui}, 
the cold atom trapped in one-dimension \cite{zhang,jiang}. The 
exotic physics of this topological state  and its application in 
non-abelian quantum computation are few of the important features of 
the Majorana modes~\cite{nayak,sau2,ali3}.

In this paper, we look for the Majorana fermion modes in a model 
Hamiltonian system which presents the physics of an interacting helical 
liquid. It generally originates in a quantum spin Hall 
system with or without Landau levels. In this system the 
counterpropagating fields with opposite spin orientations are confined 
to the edge. The spin and momentum degrees of freedom are 
coupled together in this phase. However, unlike chiral Luttinger liquid, the time 
reversal symmetry in this phase is preserved. The basic physical 
aspects of helical spin liquid are discussed in 
Refs.~\cite{sela,sarkar2,qi,qi2}.

The existence of  Majorana modes in proximity induced topological 
superconductor  is modelled using a fermionic 
model~\cite{kitaev,fu1,fu2,ali1,ali,sau,potter,ali2,sela,sarkar2,qi2}. 
The field theoretical calculation by the authors of Refs.~\cite{sela,sarkar2} shows that Majorana 
fermion modes in a helical liquid posses a higher degree of 
stability. Scattering processes between the two 
constituent fermion bands help the helical liquid to retain the 
properties by opening a gap in presence of the interaction.  The 
strong interaction may induce decoherence in the Majorana 
modes~\cite{suhas,ali2,sela,sarkar2}. However, the proximity gap 
generates the Majorana excitation~\cite{sela}. The presence of 
Majorana modes and length dependence on various parameters
are studied using renormalization group method  
by one of our co-author~\cite{sarkar2}. However, there is no 
estimation of numerical values of phase boundary in previous 
work~\cite{sela,sarkar2}. Therefore, we use the density matrix 
renormalization group (DMRG) method to calculate  accuratly the phase 
boundary of Majorana-Ising topological phase transition accurately.

\section{Brief description of the model Hamiltonian}
For the completeness of the paper here we describe in brief the
one-dimensional helical system in terms of the field operators. 
We also derive the Hamiltonian of the system in terms of the 
field operators. It is well known in literature that the low energy 
excitation in the one dimensional quantum many body system 
occurs in the region adjacent to the Fermi points. Therefore, 
one can write the fermionic field operator 
as~\cite{qi}
\begin{equation}
{\psi}_{\sigma} (x) = \frac{1}{\sqrt{L}} \left[ \sum_{-\Lambda < k - k_F < \Lambda}
e^{i k.x}\, {\psi}_{\sigma} + \sum_{-\Lambda < k + k_F < \Lambda}
e^{i k.x} \, {\psi}_{\sigma} \right], 
\end{equation}
where $\Lambda $ is the cut-off around the Fermi momentum ($k_F $).
We may consider the first term as a right mover ($ k>0 $) and the second term
as a left mover ($ k < 0 $). One can write the fermionic field with spin 
$\sigma $ as $ {\psi}_{\sigma} (x) = {\psi}_{R \sigma} (x) + {\psi}_{L \sigma} (x) $.
For the low energy elementary excitations one can write the Hamiltonian
as
\begin{eqnarray}
H_0 &=& \int \frac{dk}{2 \pi} {v_F} \left[ \left( {{\psi}_{R \uparrow}}^{\dagger} (i \partial_x)
{\psi_{R \uparrow} } - i {{\psi}_{L \downarrow}}^{\dagger} (i \partial_x)
{\psi_{L \downarrow} }\right) \right. \non \\
& & + \left. \left({{\psi}_{R \downarrow}}^{\dagger} (i \partial_x)
{\psi_{R \downarrow} } - i {{\psi}_{L \uparrow}}^{\dagger} (i \partial_x)
{\psi_{L \uparrow} }\right) \right],
\label{eq:h0}
\end{eqnarray}
where $\psi_{R\uparrow}(x)$ and $\psi_{L\downarrow}(x)$ are the field 
operators for spin up right moving and spin down left moving electrons
respectively. The terms within the parenthesis are the respective Kramer’s 
pairs. One of these Kramer’s pairs is in the upper edge and the other one is  
in the lower edge of the system. The total fermionic field of this system is,
$\psi(x) = e^{i k_F x} \psi_{R\uparrow} + e^{-i k_F x} \psi_{L\downarrow}$.
This is the simple picture of a helical liquid, where the spin is determined 
by the direction of the particle. The non-interacting part of the helical 
liquid for a single edge in terms of spinor field is
\begin{equation}
H_{01} = {\psi}^{\dagger}  (i v_F \, \partial_x   {\sigma}^z  - \mu) \, {\psi} .
\end{equation}
Now we introduce the  model Hamiltonian of the present study. The model 
Hamiltonian describes a low-dimensional quantum many body system of topological 
insulator in the proximity of s-wave superconductor and an external magnetic 
field along the edge of this system. The additional terms in the Hamiltonian is 
\begin{equation}
\delta H =  \Delta {\psi_{L \downarrow}} {\psi_{R \uparrow}} + 
B {{\psi}_{L \downarrow}}^{\dagger}{\psi_{R \uparrow}} + h.c.,
\end{equation}
where $\Delta $ is the proximity induced superconducting gap and $B$ is
the applied magnetic field along the edge of the sample. The Hamiltonian
$H_0 $ is  time reversal invariant, however the Hamiltonian $\delta H$ 
breaks the time reversal symmetry.

Now we consider the generic interaction which 
preserves time reversal symmetry. The authors of Ref.~\cite{sela,sarkar2}, have 
considered the two-particle forward and umklapp scattering. The forward scattering 
is described as 
\begin{equation}
H_{fw} = g_2  {\psi_{L \downarrow}}^{\dagger} {\psi_{L \downarrow}}
 {\psi_{R \uparrow}}^{\dagger} {\psi_{R \uparrow}} . 
\end{equation}
We write the umklapp scattering term for the half filling in a point splitted form, following 
the Wu, Bernevig and Zhang~\cite{wu}. The point splitted version can be described as a regularization 
of the theory. Therefore, the umklapp term becomes 
\begin{equation}
H_{um} = - g_u \int dx 
{{\psi}_{R \uparrow}}^{\dagger} (x) 
{{\psi}_{R \uparrow}}^{\dagger} (x + a) 
{{\psi}_{L \downarrow}}^{\dagger} (x) 
{{\psi}_{L \downarrow}}^{\dagger} (x + a) +h.c.,  
\end{equation}
where $a$ is the lattice constant. This analytical expression gives a regularized theory 
using the lattice constant $a$ as an ultraviolet cut-off. We use the first order Taylor 
series expansion of the fermionic field 
\begin{equation}
{{\psi}_{R \uparrow}}^{\dagger} (x + a) 
\sim {{\psi}_{R \uparrow}}^{\dagger} (x)  + a \partial_x 
{{\psi}_{R \uparrow}}^{\dagger} (x) .
\end{equation}
Using this expansion in the umklapp scattering term  we produce the analytical
expression for umklapp in a conventional form of the authors of Ref.~\cite{wu}.
\begin{equation}
H_{um} = g_u {{\psi}_{L \downarrow}}^{\dagger} \partial_x {{\psi}_{L \downarrow}}^{\dagger}
{{\psi}_{R \uparrow}} \partial_x {{\psi}_{R \uparrow}} + h.c
\end{equation}
Therefore the total Hamiltonian of the system is
\begin{equation}
H = H_{0} + H_{fw} + H_{um} + \delta H .
\end{equation}
Now we can write the above Hamiltonian as, 
$H_{XYZ} = \sum_i H_i $  \cite{sela}.
\begin{equation}
H_i = \sum_{\alpha} J_{\alpha} {S_i}^{\alpha} {S_{i+1}}^{\alpha}
- [ \mu + B (-1)^i ] {S_i}^z ,
\label{sela:ham}
\end{equation}
where $J_{x,y} = J \pm \Delta >0 $ and $J =v_F $ and $J_z  >0 $.

\section{Construction of Kitaev's chain for this system }
In this section we map the model Hamiltonian of the present problem to the Kitaev's 
chain. Considering  the limit $\Delta = J$, the  
Hamiltonian in Eq.~\ref{sela:ham} reduces to the transverse Ising model for $\mu =0 $ and a $\pi$ rotation 
of alternate spins~\cite{sela}. If we write the transverse Ising model Hamiltonian in terms of 
Pauli spin operators, the Hamiltonian reduces to
\begin{equation}
H =  \frac{1}{2} \sum_i ( {\Delta} {\sigma_i}^{x} {\sigma_{i+1}}^{x}+
{B} {\sigma_i}^{z} ).
\label{eq:h1}
\end{equation}
We change the sign of the magnetic field without loss of generality. One can 
write the Hamiltonian in Eq.~\ref{eq:h1} in terms of spinless fermion operators after a Jordan-Wigner
transformation. To do so, we use the relation:
$ {\sigma_i }^{z} = (2 {\psi_i}^{\dagger} {\psi_i} - 1) $,
$ {\sigma_i}^{x} {\sigma_{i+1}}^{x}  =  ({{\psi}_i }^{\dagger} - {\psi_i})
({{\psi}_{i+1} }^{\dagger} + {\psi_{i+1} }) $.
The Hamiltonian, $H_1 $ , becomes,
\begin{eqnarray}
H_{1} & = & \frac{\Delta}{2} \sum_{n} ( {\psi}^{\dagger} (n) {\psi} (n +1 ) + h.c )
+  B \sum_{n} {\psi}^{\dagger} (n) {\psi} (n) \non \\
& & + \frac{\Delta}{2} \sum_{n} ( {\psi}^{\dagger} (n) {\psi}^{\dagger} (n +1 ) + h.c ) .
\end{eqnarray}
Here ${\psi^{\dagger}}(n) ({\psi}(n)) $ is the creation(annihilation) operator 
for spinless fermion at the site $n$. After the Fourier transformation, the 
Hamiltonian, $H_1$ , reduces to,
\begin{eqnarray}
H_1  & = &  \sum_{k> 0} ( B + \Delta  \cos k)
({\psi_k}^{\dagger} {\psi_k} + {\psi_{-k}}^{\dagger} {\psi_{-k}}) \non\\
& & +  i \Delta  \sum_{k > 0} \sin k  ({\psi_k}^{\dagger} {\psi_{-k}}^{\dagger} +
{\psi_{k}} {\psi_{-k}}),
\label{eq:13}
\end{eqnarray}
where $ {\psi^{\dagger}} (k) \left(\psi (k)\right)$ is the creation (annihilation) operator 
of the spinless fermion of momentum $k$. $ H_1 $ in Eq.~\ref{eq:13} is written in terms of Kitaev's chain as
\begin{equation}
H =  \sum_{n} -{t}  ( {c_n}^{\dagger} c_{n+1} + h.c ) -
 {\mu} {c_n}^{\dagger} {c_n}
+  {|\Delta_1|} ( {c_n} {c_{n+1}} + h.c ) ,
\end{equation}
where $ t$ is the hopping matrix element, $\mu$ is the chemical potential
and $|\Delta |$ is the magnitude of the superconducting gap.  
$ {t} = \frac{\Delta}{2} $, $ {\Delta_1} = \frac{\Delta}{2} $
and ${\mu} =  B$.\\
The authors of Ref.~\cite{ronny} also study the one dimensional 
Ising model and topological order in Kitaev's chain. The authors
study the $\Delta = t$ and $\mu =0$ limit of Kitaev's chain.
They find the explicit eigenstate of the open chain in terms of
fermion operators and also show that the states as well as the energy eigen
values are equivalent to those of an Ising chain. \\
In the present study we obtain the model Hamiltonian in the form of a 
transverse Ising model for a certain regime of parameter space, and finally 
we map this model to the Kitaev's chain. Therefore, the perspective of this 
study is different from the previous study of Ref.32. \\ 

The bulk properties of Hamiltonian can be studied 
in the momentum space. One can write down the Hamiltonian 
in momentum space as.
\begin{equation}
H = (\frac{1}{2}) \sum_{k} {\psi_k}^{\dagger} H(k) {\psi_k}
\end{equation}
~~~~~~~~~~~~~~~~~~~~~~$ {H(k) } = \left (\begin{array}{cc}
      {\epsilon} (k)  & 2 {{\Delta}}^{*} (k) \\
   2 {\Delta} (k)   & - {\epsilon} (k)
        \end{array} \right ) $
where,
$ {\epsilon} (k) = -2 t \cos k - {\mu}, $
and $ {\Delta}(k) = - i {\Delta} \sin k $.
These Hamiltonians correspond to the p-wave superconducting
phase, one can understand this in the following way. One can also 
write down the above Hamiltonian in Bogoliubov energy spectrum,
\begin{equation}
H_{1} = \sum_{k}  { E(k)} {c_k}^{\dagger} {c_k}
\end{equation}
Here ${E_k} (= \sqrt{ {({\epsilon_k } - \mu )}^2 + {{{\Delta}_k}^2 } } ) $
is the energy spectrum in bulk and ${c_k}^{\dagger} $ and $c_k $
are the Bogoliubov quasiparticles operators.
It is well known in the literature that the Kitaev's chain consists
of topological properties. Here we discuss it very
briefly following the Refs.~\cite{berni} and~\cite{kitaev}. \\ 
One can express the Dirac Hamiltonian of the system in terms Majorana fermion modes
which are linear combination of fermionic operators. \\
$ {c_N} = \frac{1}{2} ( a_{2N-1} + i a_{2N} )$ 
and 
$ {c_N}^{\dagger} = \frac{1}{2} ( a_{2N-1} - i a_{2N} )$ and the anticommutation
relation between the Majorana fermion modes is 
$ \{ {a_N}^{\dagger}, a_{N'} \} =2 \delta_{N N'}$. The non-topological
phase of the Kitaev's chain appears for the following limit.\\
(A). $\mu < 0 $ and $ |\Delta | = t =0 $,\\
$ H = ( \frac{-i \mu}{2} ) \sum_N ( a_{2N -1} a_{2N} ) $. 
For the 
present problem the above Hamiltonian becomes as \\  
$ H = ( \frac{-i B}{2} ) \sum_N ( a_{2N -1} a_{2N} ) $. \\ 
In  this phase Majorana operators couple on each site and there is no intersite 
coupling.\\
(B). The topological phase $ |{\Delta}| = t > 0$ and $\mu =0 $:
The Kitaev's chain reduces to \\
$ H = (it) \sum_N { a_{2N} a_{2N -1}} $. \\
For the present problem the above Hamiltonian is reduced to \\
$ H = (i \frac{\Delta}{2}) \sum_N { a_{2N} a_{2N -1}} $. \\
It is clear from this analytical relation that the intersite
Majorana fermions are coupled in the lattice however, $a_1 $ and 
$a_{2N}$ are not coupled to the rest of the chain and they 
are unpaired. For this case, zero modes are localized at the
ends of the chain.\\

In present numerical studies the topological quantum phase transitions 
are studied for all the regime of parameters. 
Before going to the numerical section, let us discuss the condition for appearance
of Majorana fermion edge mode briefly: In a nanowire or at the 
edge of topological insulator where the helical spin liquid appears, the zero mode Majorana
edge state appears as the particle-hole bound state at both ends
of the wire or edge with localization length ($\xi \sim \frac{v}{\Delta} $)~\cite{sarkar2,suhas,dip}.
The overlap of the Majorana wave functions is proportional to $e^{-N/\xi }$, 
$N$ is the length of the system. The existence of the Majorana fermion 
zero mode can also be characterised by exponential decay of lowest 
excitation gap with system size. There are many numerical studies on the  Kitaev or interacting 
Majorana chain, topological superconducting wire and others~\cite{gergs,affleck_dmrg,ali2,haldane_dmrg,satoshi}.


\section{DMRG study based results for Majorana-Ising transition 
and Majorana localization length  }
In this section, we numerically solve the Hamiltonian mentioned in Eq.~\ref{sela:ham} using the DMRG method.
This method is a state of the art numerical technique for 1D system, and it is based 
on the systematic truncation of irrelevant degrees of freedom in the Hilbert space
\cite{white-prl92,*white-prb93}. This numerical method is best suited to calculate accurate 
ground state (GS)  and a few low lying energy excited states of 
strongly interacting quantum systems. For ladders  and long range interaction systems 
the DMRG is further improved by modifying conventional DMRG method to  solve  
chain with periodic boundary condition~\cite{dd2016a}, zigzag chains~\cite{mk2010}, 
the Y-junction systems~\cite{mk2016} and Bethe lattice~\cite{mk2012} etc. 
The left and right block symmetry of DMRG algorithm for a XYZ model of a spin-1/2 chain  
in a staggered magnetic field (Eq.~\ref{sela:ham}) is broken. 
Therefore, we use conventional unsymmetrized DMRG algorithm, where 
the left and right block are unequal in general. This model does not conserve the total  $S^z$, 
therefore superblock dimension is large. We keep $m \sim 400$ eigenvectors corresponding to 
the highest eigenvalues of the density matrix to  maintain
excellent accuracy of eigenvalues and eigenvectors of the superblock. The truncation 
error of density matrix eigenvalues is less than $10^{-11}$. The energy 
convergence is better than $0.001\%$ after five finite DMRG sweeps.  
We carry out the  DMRG calculation for various parameter regimes of the system up to $N = 200$ with 
open boundary condition (OBC). 

\begin{figure}
\centering
\includegraphics[scale=0.4,angle=0]{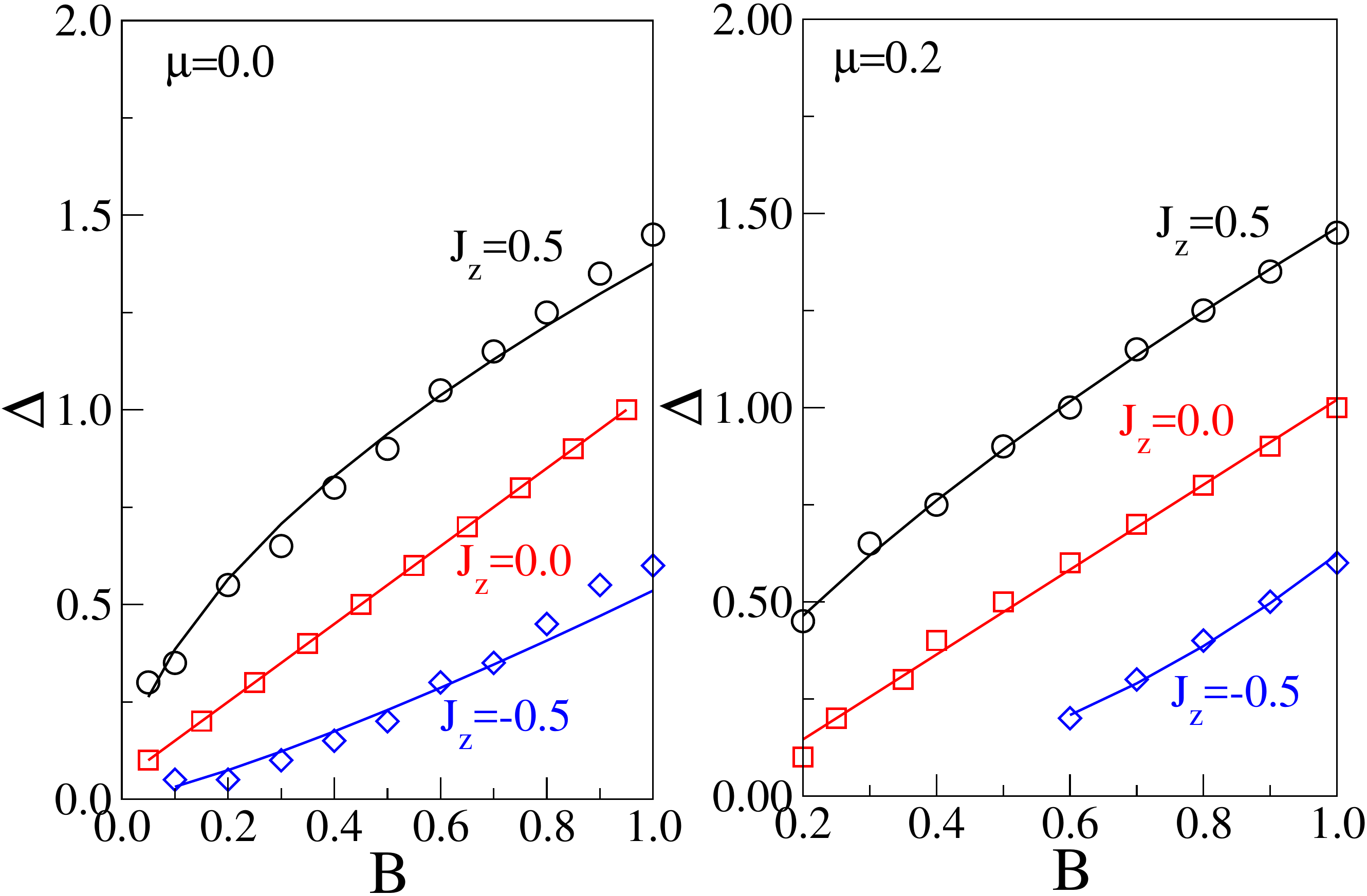}
\caption{ (Colour online.)
Majorana-Ising (MI) phase transition in parameter space of $\Delta$ and $B$ for 
$\mu=0.0$ (left panel) and $\mu=0.2$(right panel) for $N = 100$ 
sites, and  $J_z= 0$ and $\pm 0.5$. The phase transition points 
follow power law and the exponents are 0.55, 1.0, 1.23 (left panel) and 
0.71, 1.0, 2.14 (right panel) for $J_z = 0.5$, 0.0 and $-0.5$ respectively.}
\label{Fig2}
\end{figure}
\begin{figure}
\centering
\includegraphics[scale=1.2,angle=0]{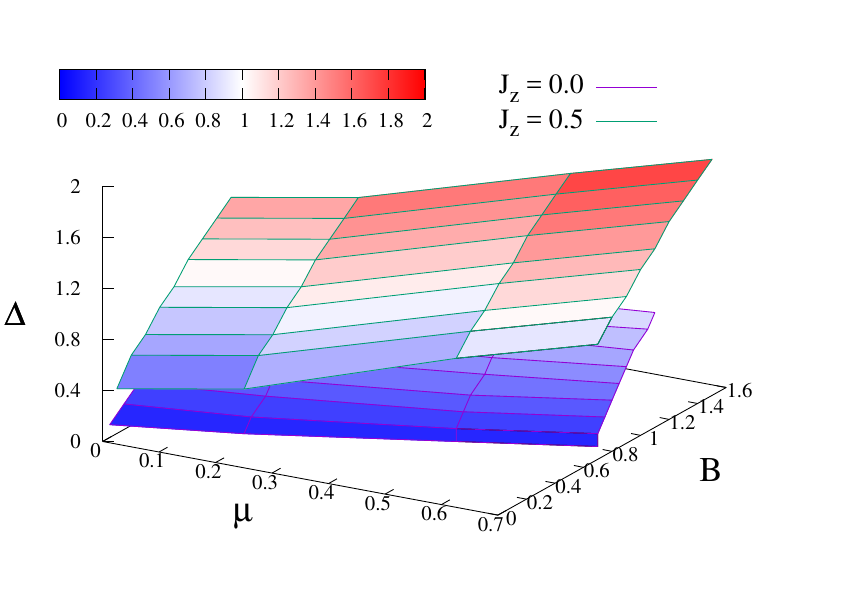}
\caption{ (Colour online.)
MI phase transition surface in parameter space of $B$, $\mu$, and
$\Delta$  for a chain of $N = 100$ sites. Upper surface along $\Delta$-axis is for 
$J_z=0.5$ and lower surface is for  $J_z=0.0$}
\label{Fig3}
\end{figure}
\begin{figure}
\centering
\includegraphics[scale=0.4,angle=0]{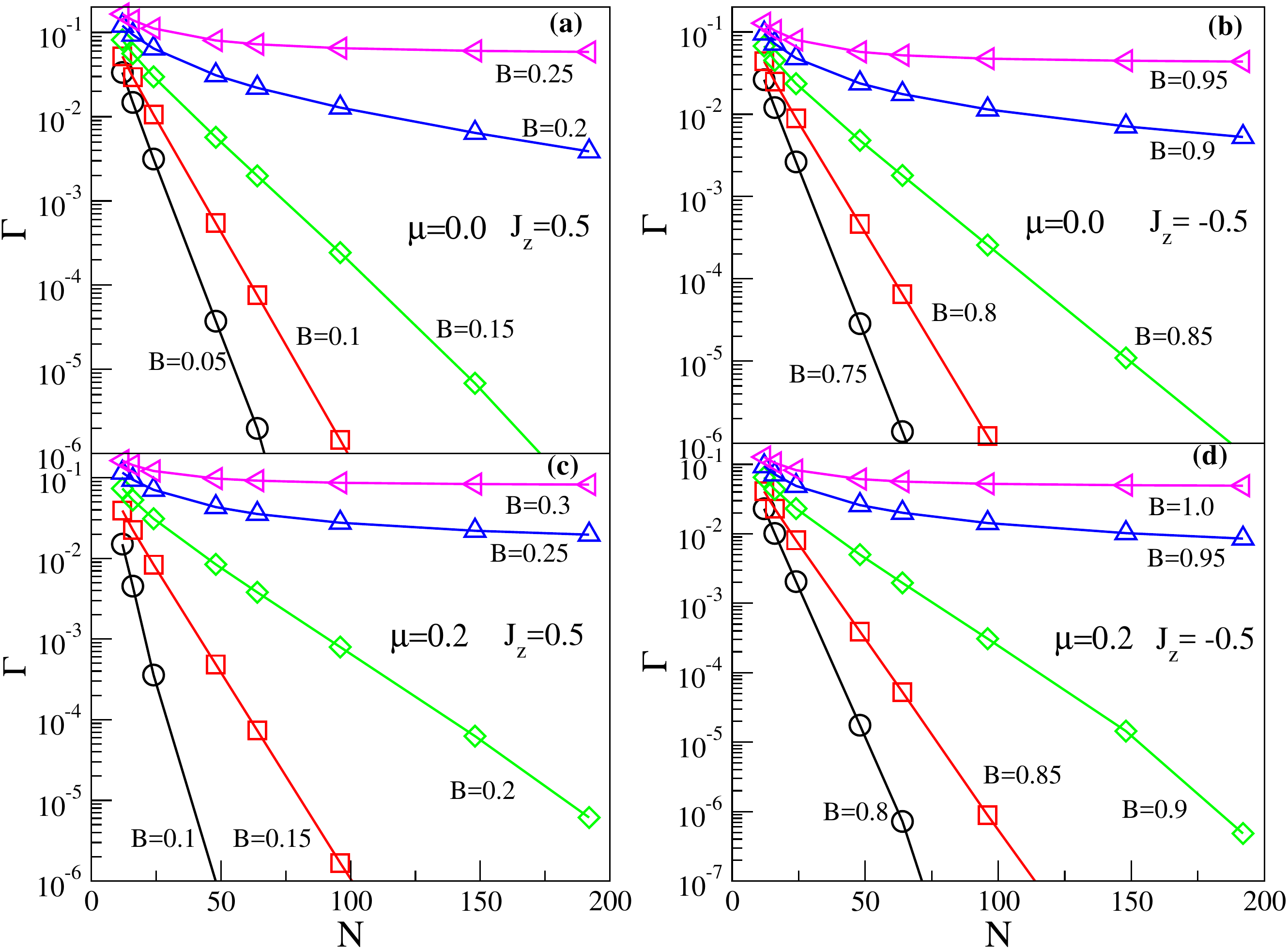}
\caption{ (Colour online.) 
The lowest excitation gap $\Gamma$ vs. $N$ for $\Delta = 0.5$ and 
various values of $B$ is shown for $\mu = 0$, (a) $J_z = 0.5$,  (b) $J_z = -0.5$ and 
$\mu = 0.2$, (c) $J_z = 0.5$, (d) $J_z = -0.5$. $\Gamma$ 
varies exponentially with $N$ in the  Majorana regime of the parameter space.}
\label{Fig4}
\end{figure}

The DMRG method is used to get a better understanding of phase transition and accurate 
phase boundary of the Majorana-Ising topological quantum phase transition 
in various parameter regime. In this section we show the Majorana-Ising phase boundary 
in Fig.~\ref{Fig2}. And based on these boundaries we construct the 3D phase 
diagram in $\Delta$-$\mu$-$B$ parameter space (in Fig.~\ref{Fig3}). 
We show the lowest excitation gap decays as a function of system size $N$ in Fig.~\ref{Fig4}.
The Majorana edge mode survives at the edge of system if a system of size $N$ holds
the condition $\frac{N \Delta}{v} \gg 1$ where $\Delta$ and $v$ are superconducting
gap and velocity of collective modes of the system. If the localization
length is defined as $\xi \sim \frac{v}{\Delta}$ \cite{ali,sarkar2,suhas,dip},
then the condition is reduced to $\frac{N}{\xi} \gg 1$.
We calculate $\Delta_c$ as a function of $N$ to calculate $\xi$ as $\Delta_c$ conserves 
exponentially and show that $\frac{N}{\xi} > 4$. At the end of this section 
we explain the origin of the excitation in different phases using a local excitation energy gap $\Gamma_i$.

In Fig.~\ref{Fig2}, the left and right panel is for the $\mu =0$ and $\mu =0.2$ respectively.  
We consider $J_z = 0.5$ for repulsive interaction, $J_z=-0.5$ for attractive interaction, and $J_z = 0$ for 
non-interacting limit.  
For $J_z =0$, in Fig.~\ref{Fig2}, behaviour is almost same for two sets of chemical potentials, and  the behaviour 
is linear similar to the Fig. 1 of Ref.~\cite{sarkar2}. In the presence of repulsive interaction ($J_z =0.5$), 
the phase boundary of this transition follows the power law variation with 
positive exponent less than one, but the phase transition line is shifted 
towards the higher values of $\Delta$ for $\mu =0.2$.   For the attractive 
interaction ($J_z =-0.5 $), the values of powers are higher than one, and it is 
consistent with the quantum field theoretical study. We notice that the 
power law exponent increases with $\mu > 0$.
The explicit $J_z$ and chemical potential dependence are absent in 
Fig. 1 of Ref.~\cite{sarkar2}. We note that the repulsive interaction shifts 
the phase boundary to the higher values of $\Delta $, but the attractive interaction 
shifts the phase boundary  to  the lower values of $\Delta$.

In Fig.~\ref{Fig3}, we present the three dimensional plot, which depicts the Majorana-Ising
phase transition explicitly. The phase digram in terms of $\Delta$, $\mu$ and $B$ 
is not possible from the study of anomalous scaling of dimensional analysis of Ref.~\cite{sarkar2}.
In this figure, we use a  wider range
of $\Delta$, $\mu$ and $B$. We observe the sharp difference of phase boundary
between the Majorana and the Ising phase for both interacting and non-interacting case. 


Fig.~\ref{Fig4} consists of four panels (a,b,c,d), the first two panels (a and b) are for
$\mu =0$ and the other two (c and d) are for the $\mu =0.2$ respectively. Here we
present the results for the lowest excitation gap $\Gamma$ in different phases. 
At first, we present the results for $\mu =0$ for both repulsive ($J_z =0.5$, panel a) 
and attractive ($J_z =-0.5$, panel b) interaction for different values of B. In this 
study, we present $\Gamma$ as a function of the system size and also show that if the 
$\Gamma$ is the Majorana mode excitation, it decays exponentially with system size $N$, 
it follows power law otherwise. The exponential decay of the lowest excitation gap
is very similar to the existence of edge states in spin-1 Heisenberg antiferromagnetic
chains~\cite{haldane83,*dd2016b}. It reveals from our study that for attractive 
interactions the system shows the existence of Majorana fermion mode for the larger 
values of $B$ ($=0.85 $) for the larger length scale,  but in the presence of 
the repulsive interaction the elementary excitation 
gap becomes finite for small values of external magnetic field ($B=0.25 $). 
We present the results for finite $\mu (=0.2)$ in the lower panels. In panel c and d 
results are shown for the repulsive interaction $(J_z =0.5 )$  and 
the attractive interaction $(J_z = -0.5)$. We note that for finite $\mu $ 
the excitations gap shows the gapless excitations for higher values of $B$ compared to
the $\mu =0 $ case whether $J_z$ is positive or negative. In the above parameter regime,  
perturbative  RG methods can not be applied and it is extremely difficult to calculate 
the excitation gap with this analytical method. 

\begin{figure}
\centering
\includegraphics[scale=0.4,angle=0]{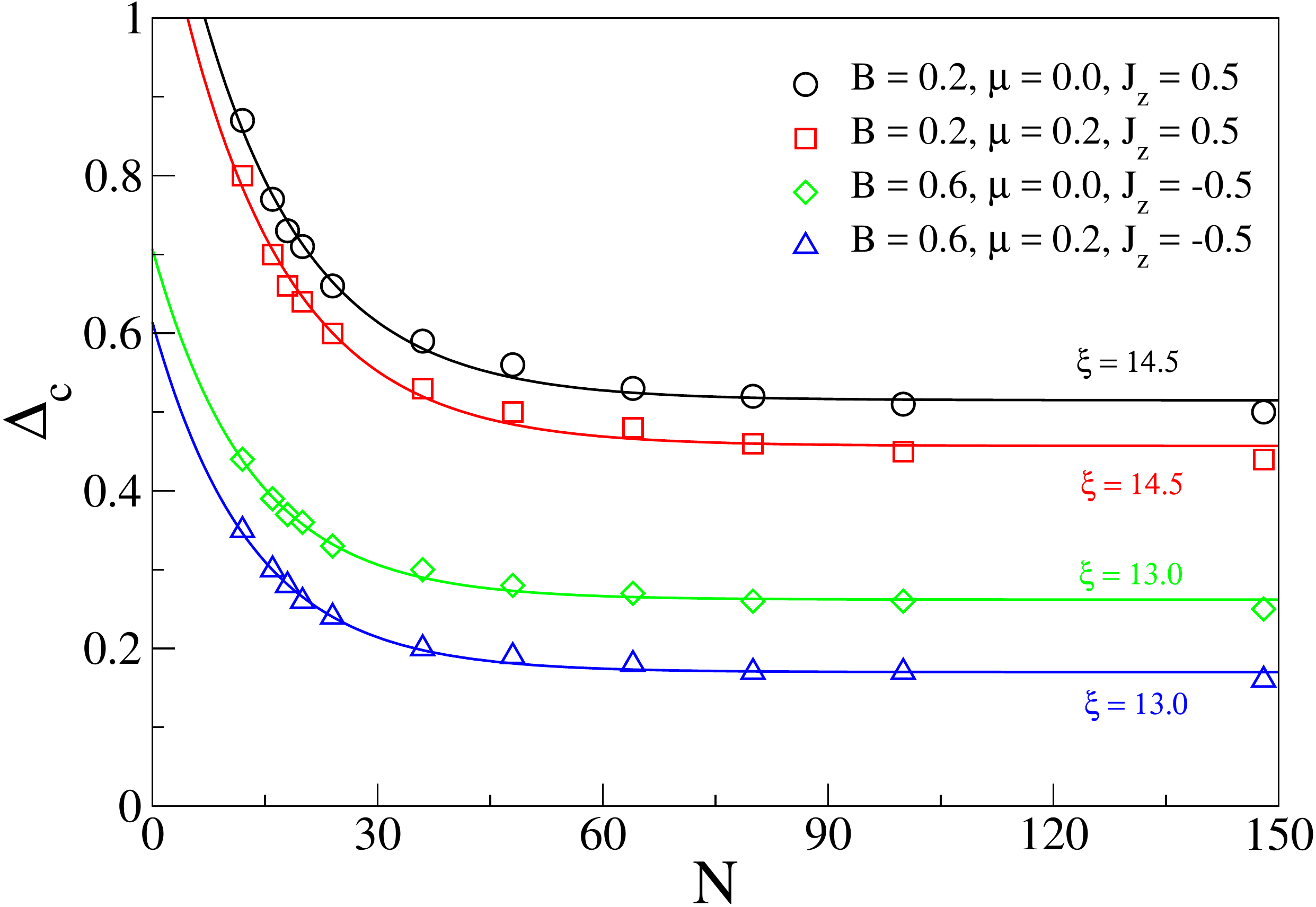}
\caption{ (Colour online.) 
Threshold  $\Delta_c$ for  $B = 0.2$ and 0.6, $\mu = 0.0$ and 0.2, and
$J_z= \pm 0.5$ with  system sizes. The solid lines are the exponential fit: 
$\Delta_c = \Delta_0 + A \exp \left(-\frac{N}{\xi}\right)$.
The fitted values of $\xi$ are given on the curves.}
\label{Fig5}
\end{figure}

 In Fig.~\ref{Fig5}, the variation of the critical values of the 
superconducting proximity ($\Delta_c$) with the system size ($N$) 
is shown for two chemical potentials ($\mu = 0, \, 0.2$) with $B = 0.2$ for $J_z = 0.5$
and $B = 0.6$ for $J_z = -0.5$. For all the cases $\Delta_c$ decays
exponentially with $N$ to a constant value $\Delta=\Delta_0$. The values 
of $\Delta_0$ depend on the set of the parameters considered.
We fit the calculated $\Delta_c$ with the equation  $\Delta_c = \Delta_0 + A \exp \left(-\frac{N}{\xi}\right)$ 
where $\xi$ is the  localization length for the Majorana mode for a given parameter.  The average 
value of $\xi$ is approximately  $15$ which is much smaller than the system size $N=100$. 
As mentioned in the second paragraph of this section, the condition for existence of 
Majorana mode is $\frac{N}{\xi} \gg 1$. The results for few representative values 
of $B$, $\mu$ and $J_z$ are shown in Fig.~\ref{Fig5}. $\Delta_0$ depends on the 
parameters $\mu$ and $J_z$ for a fixed value of $B$. For a typical value of 
$\mu$ and $J_z$ the behaviour of $\Delta_c$ curve is similar. However, $\Delta_0$ varies linearly with $B$.

\begin{figure}
\centering
\includegraphics[scale=0.4,angle=0]{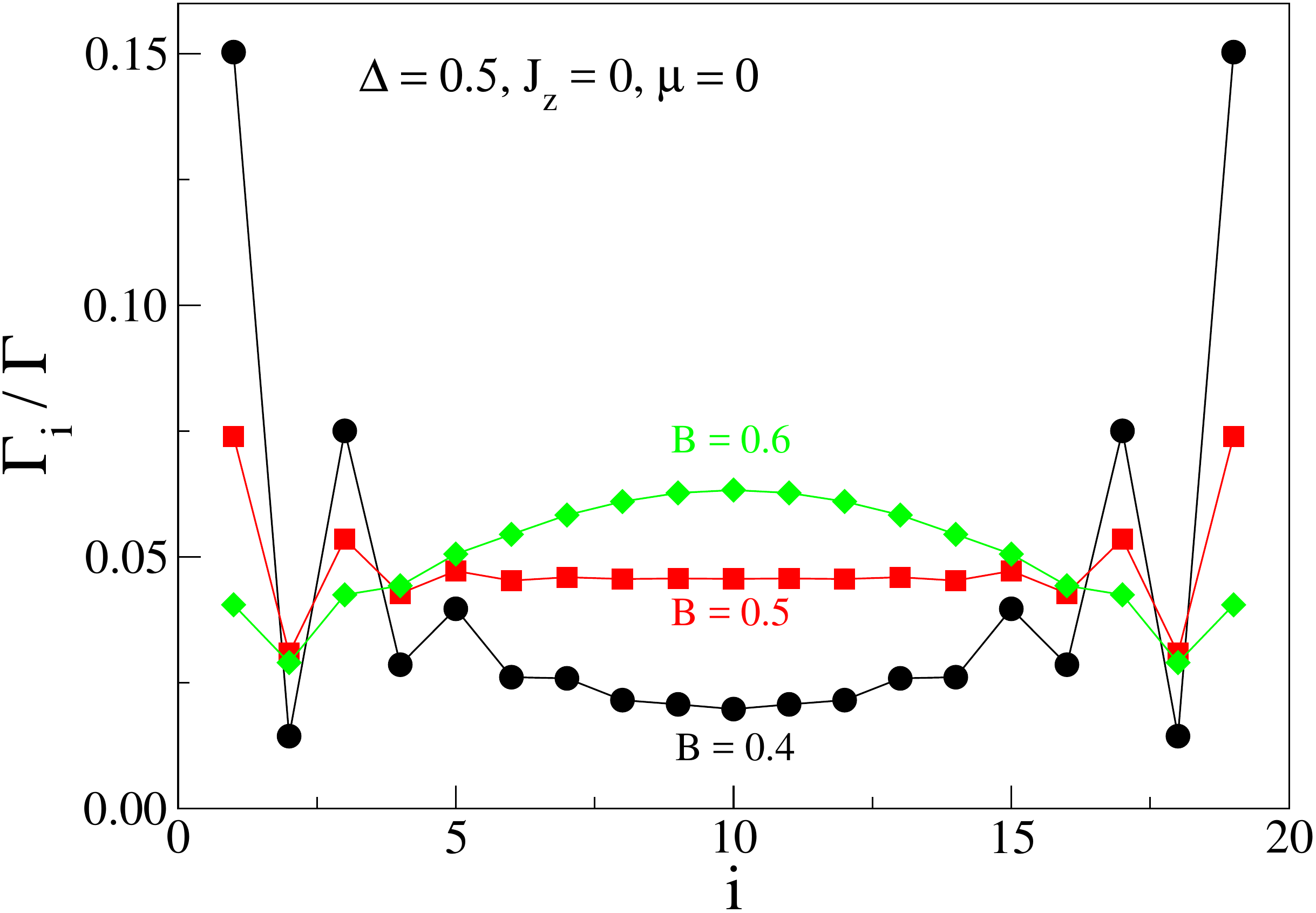}
\caption{ (Colour online.)  The lowest excitation energy gap 
$\Gamma$ at each bond of a chain of $N = 20$ spins at 
$\Delta = 0.5$ and $J_z = \mu = 0$ for $B = 0.4$ (Majorana regime), 
$B = 0.5$ (MI transition point) and $B = 0.6$ (Ising regime).}
\label{Fig7}
\end{figure}

To show the existence of Majorana modes more explicitly we calculate the
contribution of local bond and site energy $\epsilon_i$ to the lowest excitation energy gap $\Gamma$.
The local excitation energy gap $\Gamma_i$ is the difference between $\epsilon_i$ in the 
GS and the lowest excited state:
\begin{equation}
\Gamma_i = \epsilon_i^1 - \epsilon_i^0 = 
\langle \psi_1|H_i|\psi_1 \rangle - \langle \psi_0|H_i|\psi_0 \rangle,
\label{eq23} 
\end{equation}
where $H_i$ is defined in Eq.~\ref{sela:ham}, and $|\psi_0 \rangle$ and $|\psi_1 \rangle$ 
are the GS and  the first excited state, and $\epsilon_i^0$ and 
$\epsilon_i^1$ are the local energies in the GS and the first excited 
state respectively. We normalize this local excitation gap $\Gamma_i$ 
by the total energy gap $\Gamma$ such that the sum of the ratios $\frac{\Gamma_i}{\Gamma}$ 
is unity. In Fig.~\ref{Fig7}, $\frac{\Gamma_i}{\Gamma}$ is shown where, the $\mu = J_z = 0$ 
case is considered. Three parameter regimes near the phase boundary
are shown in Fig.~\ref{Fig7} for $\Delta = 0.5$ and $B$ = 0.6, 0.5 and 0.4. 
It is clear from the curves that the contribution to the excitations 
comes mainly from the edge of the chain in the Majorana phase.
Whereas, in Ising limit excitation energy contribution comes mainly from the bulk.
The critical point $\Delta = B$ shows the intermediate behaviour.

\section{Conclusion}
We have studied the Majorana-Ising quantum phase transition of helical
spin liquid system using the DMRG method. We have calculated
the Majorana localization length in various parameter regimes. The 
exponential decay of the gap with N has been shown. We have also showed 
that the major contribution to the lowest excitation gap in the topological 
state is from the edge, whereas it comes from the bulk in the Ising phase.

\begin{acknowledgments}
SS thanks the DST (SERB, SR/S2/LOP-07/2012) 
fund and also the library of RRI for extensive support. MK thanks 
DST for a Ramanujan Fellowship SR/S2/RJN-69/2012 and funding computation 
facility through SNB/MK/14-15/137.    
\end{acknowledgments}

\bibliography{ref_MI}

\end{document}